\documentclass[a4paper,11pt]{article}
\usepackage{float}
\usepackage{jcappub}
\usepackage{lineno}

\usepackage{bm}
\usepackage{mathrsfs}
\usepackage[T1]{fontenc}
\usepackage[normalem]{ulem}
\usepackage{graphicx}
\usepackage[caption=false]{subfig}
\usepackage{booktabs}
\usepackage{dcolumn}
\usepackage{multirow}
\usepackage{hhline}
\usepackage{placeins}
\usepackage[british]{babel}
\usepackage{xcolor}
\usepackage{xparse}
\usepackage{orcidlink}
\usepackage[bb=boondox]{mathalpha}
\usepackage{enumitem}
\usepackage{makecell}
\usepackage{threeparttable}

\graphicspath{{figs/}}

\allowdisplaybreaks[1]

\title{\boldmath A No-Go Theorem for Type-3 New General Relativity on $\Lambda$CDM background}

\author[a]{Shivam Kumar Mishra\,\orcidlink{0009-0006-4754-4103}}
\author[b]{B. Mishra\,\orcidlink{0000-0001-5527-3565}}

\affiliation[a,b]{Department of Mathematics,\\
Birla Institute of Technology and Science-Pilani,\\
Hyderabad Campus, Jawahar Nagar, Kapra Mandal,\\
Medchal District, Telangana 500078, India.}

\emailAdd{shivamkumarmishra.mt@gmail.com}
\emailAdd{bivu@hyderabad.bits-pilani.ac.in}

\abstract{
We investigate whether Type-3 New General Relativity can alleviate the cosmic growth tension. On a $\Lambda$CDM background, we derive the scalar perturbation equations and find the effective gravitational coupling to exceed Newton's constant throughout the parameter range compatible with ghost-free and background-hierarchy conditions. By confronting the theoretical model with growth-rate measurements from major galaxy redshift surveys, including WiggleZ, SDSS-IV eBOSS, BOSS, and VIPERS, we obtain constraints on the effective coupling that are consistent with the $\Lambda$CDM value. Consequently, the conditions for stability and tension reduction are mutually exclusive.
}

\begin{document}
\maketitle
\raggedbottom

\section{Introduction}\label{sec:intro}
The standard cosmological model faces persistent tensions. Local measurements of the Hubble constant disagree with the cosmic microwave background at around $5\sigma$ level~\cite{Riess:2019cxk,Planck:2018vyg,DiValentino:2020zio,CosmoVerseNetwork:2025alb}. Weak-lensing and galaxy-clustering surveys prefer a lower amplitude of matter fluctuations than predicted~\cite{ACT:2023kun,DiValentino:2020vvd}.~These tensions motivate modifications of gravity and of the dark sector. However, the interpretation of such tensions is not straightforward. Recent work shows that the preference for dynamical dark energy is not a model-independent statement, but depends on the assumed early-universe physics~\cite{Schoneberg:2026buf}. Any claim of a new cosmic ingredient must therefore be assessed together with the early cosmology it assumes.

Teleparallel gravity offers a geometric alternative to general relativity, describing gravity by torsion rather than curvature. New General Relativity (NGR) is the most general parity-preserving theory quadratic in torsion~\cite{Einstein1928,Hayashi:1979qx}. It contains three free parameters, which yield nine distinct subclasses. The presence of these additional degrees of freedom motivates NGR as a candidate framework for cosmology. However, the viability of these subclasses depends critically on their constraint structure. The Dirac-Bergmann analysis reveals that Type-6 is equivalent to GR, Type-9 lacks tensor modes, and Type-1 suffers from ghost instabilities in the literature~\cite{Blixt:2018znp,Tomonari:2024ybs,Bahamonde:2024zkb}. 

Type-3 NGR is the only subclass that correctly captures all degrees of freedom of the full theory without ghost instabilities~\cite{Golovnev:2023jnc,Tomonari:2025kgb}. This makes it the most promising candidate for cosmological applications. To ensure that cosmological perturbation theory remains valid, however, the theory must remain free of strong coupling issues. There are two notions of strong coupling in gravity. The conventional one refers to derivative interactions overwhelming the kinetic term. The second, more relevant for expanding backgrounds, occurs when the dominant contribution of a perturbation switches from its kinetic term to the background terms. It is referred as the perturbativity/background-hierarchy bound~\cite{JimenezCano:2021rlu}. For Type-3 NGR, ghost-freedom requires $c_2>0$ (tensor modes) and $c_3>2c_2$ (scalar and vector modes), and the hierarchy requirement adds $c_3<4c_2$ together, $2c_2<c_3<4c_2$ ~\cite{Tomonari:2026lir}. Any cosmological test of Type-3 NGR must respect these conditions.

In this work, we ask whether Type-3 NGR can reduce the $S_8/f\sigma_8$ tension. We derive the scalar perturbation equations, obtain the M\'esz\'aros equation, and confront the predicted growth history with $f\sigma_8$ measurements. The analytic analysis is restricted to the range allowed by the no-ghost and hierarchy bounds, and a wide-prior run is shown only as a diagnostic.

The paper is organized as follows. Section~\ref{NGR:intro} reviews NGR and derives the background equations in the presence of a cosmological constant. Section~\ref{sec:pertmatter} presents the scalar perturbation equations, the antisymmetric constraint, and their reduction to the M\'esz\'aros equation under the sub-horizon and
quasi-static approximations, from which the effective gravitational coupling is extracted. Section~\ref{sec:fs8} confronts the predicted growth history with $f\sigma_8$ measurements using an MCMC analysis. Section~\ref{sec:conclusions} summarizes our conclusions. We adopt units with $c=\hbar=1$ and metric signature $(-,+,+,+)$.

\section{New General Relativity}\label{NGR:intro}
New General Relativity (NGR) constitutes a three-parameter extension of the Teleparallel Equivalent of General Relativity (TEGR). The latter belongs to a specific subclass of metric-affine gauge theories of gravity (MAG) and reproduces the dynamics of General Relativity (GR) up to a total boundary term in the action functional. Comprehensive reviews of these frameworks can be found in Refs.~\cite{Hehl:1994ue,Bahamonde:2021gfp}.

In both TEGR and NGR, the affine connection is constructed from the tetrad \(e_{a}{}^{\mu}\) and the co-tetrad \(e^{a}{}_{\mu}\) according to
\begin{equation}
\Gamma^{\rho}{}_{\mu\nu}
=
e_{a}{}^{\rho}\,\partial_{\mu} e^{a}{}_{\nu},
\label{Weitzenboeck connection}
\end{equation}
which gives rise to the torsion tensor
\begin{equation}
T^{\rho}{}_{\mu\nu}
=
2\Gamma^{\rho}{}_{[\mu\nu]}
=
e_{a}{}^{\rho}
\left(
\partial_{\mu} e^{a}{}_{\nu}
-
\partial_{\nu} e^{a}{}_{\mu}
\right).
\label{torsion in W-gauge}
\end{equation}
More generally, the affine connection can be written in the form
\begin{equation}
\Gamma^{\rho}{}_{\mu\nu}
=
e_{a}{}^{\rho}\,\partial_{\mu} e^{a}{}_{\nu}
+
\omega^{a}{}_{b\mu}\, e^{b}{}_{\nu}\, e_{a}{}^{\rho},
\label{affine connection}
\end{equation}
where \(\omega^{a}{}_{b\mu}\) denotes the spin connection. Imposing the Weitzenböck gauge,
\begin{equation}
\omega^{a}{}_{b\mu} = 0,
\end{equation}
reduces the connection to Eq.~\eqref{Weitzenboeck connection}, which is therefore referred to as the Weitzenböck connection.

The gravitational Lagrangian of NGR is constructed from all inequivalent quadratic invariants formed from the torsion tensor~\cite{Hayashi:1979qx},
    \begin{equation}
\begin{split}
L_{\rm NGR}
:=
2\kappa^{2}e^{-1}\mathcal{L}_{\rm NGR}
&=
c_{1}\,
T^{\mu\nu\rho}T_{\mu\nu\rho}
+
c_{2}\,
T^{\mu\nu\rho}T_{\rho\mu\nu}
+
c_{3}\,
T^{\mu}{}_{\mu\rho}
T^{\nu}{}_{\nu}{}^{\rho}
\\[4pt]
&=
c_{1}\, g_{\mu\sigma} g^{\nu\lambda} g^{\rho\kappa} T^{\mu}{}_{\lambda\kappa} T^{\sigma}{}_{\nu\rho}
+
c_{2}\, g^{\nu\lambda} T^{\mu}{}_{\lambda\rho} T^{\rho}{}_{\mu\nu}
+
c_{3}\, g^{\rho\nu} T^{\mu}{}_{\mu\rho} T^{\lambda}{}_{\lambda\nu},
\end{split}
\label{Lagrangian of NGR}
\end{equation}
where \(e\) denotes the determinant of the co-tetrad and \(c_{1}\), \(c_{2}\), and \(c_{3}\) are arbitrary real constants. TEGR is obtained for the parameter choice
\begin{equation}
c_{1}=-\frac14,\qquad
c_{2}=\frac12,\qquad
c_{3}=1,\label{tegrchoice}
\end{equation}
for which the theory becomes dynamically equivalent to GR, differing only by a boundary term in the action, despite being formulated on a distinct underlying teleparallel spacetime geometry.

The field equations are derived by varying the total action with respect to the tetrad:
\begin{align}
\mathbb{E}_a{}^{\mu}
&:= 
e^{-1}
\partial_{\mu}
\left(
e\, e_{a}{}^{\rho} S_{\rho}{}^{\mu\nu}
\right)
+
e_{a}{}^{\lambda}
T^{\rho}{}_{\mu\lambda}
S_{\rho}{}^{\mu\nu}
-
\frac14
e_{a}{}^{\nu}
L_{\rm NGR}=\frac{\kappa^{2}}2\, e_{a}{}^{\rho}\, \Theta_{\rho}{}^{\nu},\label{field equation of NGR}
\\[4pt]
\text{where }\,\,S_{\rho}{}^{\mu\nu}
&=
c_{1}\, T_{\rho}{}^{\mu\nu}
+
c_{2}\, T^{[\mu\nu]}{}_{\rho}
+
c_{3}\, \delta_{\rho}{}^{[\mu} T^{|\kappa|}{}_{\kappa}{}^{\nu]}\,.\nonumber
\end{align}
Here,
\begin{equation}
e_{a}{}^{\rho}\,\Theta_{\rho}{}^{\nu}
=
e^{-1}
\frac{\delta\mathcal{L}_{\rm matter}}{\delta e^{a}{}_{\nu}}
\end{equation}
denotes the matter energy-momentum tensor projected onto the tetrad frame and $\kappa^{2}=8\pi G$ is the gravitational coupling constant.

The various realizations of NGR are classified by the parameter space \((c_{1},c_{2},c_{3})\). In particular, the so-called Type-3 subclass is defined by imposing the linear constraint~\cite{Blixt:2018znp}
\begin{equation}
2c_{1} + c_{2} = 0.
\label{Type 3 parameter condition}
\end{equation}
In modified teleparallel gravity, the field equations are decomposed into symmetric and antisymmetric parts, with the latter describing frame-dependent effects and coinciding with the equations obtained via variation with respect to spin connection in the covariant formulation~\cite{Golovnev:2017}. We get equations in spacetime indices from \eqref{field equation of NGR} as $ \mathbb{E}_{\mu\nu}=g_{\nu\rho}\,e^a\,_{\mu}\,\mathbb{E}_a\,^{\rho}$ and split the symmetric and antisymmetric parts as following
\begin{equation}
\mathbb{E}_{\mu\nu}
=
\mathbb{E}_{(\mu\nu)}
+
\mathbb{E}_{[\mu\nu]}.
\label{symmetric split}
\end{equation}
To ensure the symmetry of the energy-momentum tensor and maintain local Lorentz invariance, the antisymmetric part must vanish,
\begin{equation}
\mathbb{E}_{[\mu\nu]}=0.
\label{antisymmetric}
\end{equation}
\subsection{Background Cosmology with a cosmological constant}

To describe the homogeneous and isotropic expanding Universe, we adopt the Friedmann--Lemaître--Robertson--Walker (FLRW) line element,
\begin{equation}
ds^{2} = - dt^{2} + a^{2}(t)\,\delta_{\hat{i}\hat{j}}\,dx^{\hat{i}}dx^{\hat{j}},
\label{eq_25}
\end{equation}
where \(a(t)\) is the scale factor, and hatted indices label the spatial coordinates. This metric is obtained from the diagonal co-tetrad
\begin{equation}
\left[e^{a}{}_{\mu}\right]
=
\mathrm{diag}\!\left(1,a(t),a(t),a(t)\right)\,.
\label{eq_26}
\end{equation}
with determinant
\[
e=a^{3}(t).
\]
The energy-momentum tensor of a perfect fluid with energy density $\rho$ and pressure $p$ is
\begin{equation}
\Theta_{\mu\nu}
=
(\rho+p)\mathbb{n}_{\mu}\mathbb{n}_{\nu}
+
p\,g_{\mu\nu},
\end{equation}
where $\mathbb{n}^\mu\,\mathbb{n}_\mu=-1$. A cosmological constant can be interpreted as a perfect fluid with
\begin{equation}
\rho_{\Lambda}=\frac{\Lambda}{8\pi G},
\qquad
p_{\Lambda}=-\rho_{\Lambda},
\end{equation}
so that the total energy-momentum tensor becomes
\begin{equation}
\Theta_{\mu\nu}^{\rm (tot)}
=
(\rho+p)\mathbb{n}_{\mu}\mathbb{n}_{\nu}
+
\left(p-\rho_{\Lambda}\right)g_{\mu\nu}\,.
\end{equation} 

The background equations \eqref{symmetric split} are given by 
\begin{subequations}\label{background}
\begin{align}
3\left(2c_1-c_2+3c_3\right)H^2
&=
2\left(\Lambda+\kappa^2\rho\right),
\label{bgtt}
\\
\left(2c_1-c_2+3c_3\right)
\left(3H^2+2\dot{H}\right)
&=
2\left(\Lambda-\kappa^2 p\right).
\label{bgii}
\end{align}

\end{subequations}
The antisymmetric part \eqref{antisymmetric} is an identity at the background. Since the matter Lagrangian is invariant under diffeomorphisms, the energy-momentum tensor is conserved with respect to the Levi-Civita connection, resulting in the following equation at the background
\begin{equation}
\bar{\nabla}_{\nu}\Theta_{0}\,^{\nu}\equiv \dot{\rho}+3H(\rho+p)=0 \label{matterconservationbackground}
\end{equation}
where the spatial components vanish identically at the background.
\subsection{Scalar Perturbations in NGR}

Scalar perturbations, originating from fluctuations in density and pressure, govern the development of large-scale structure and induce temperature anisotropies in the CMB, thereby providing direct information about the physics of the early Universe. To investigate the growth of structure, we work with the linearly perturbed metric expressed in the Longitudinal gauge as 

\begin{equation}
{\left[\delta g_{\mu\nu}\right]} =
\left[
\begin{array}{cc}
-2\Phi & 0\\
0 & 2a^{2}\Psi\,\delta_{\hat{i}\hat{j}}
\end{array}
\right]\,.
\label{pertmet}
\end{equation}

To prevent issues with the integrated Sachs--Wolfe effect \cite{Zheng_2011}, the metric perturbation is determined by the choice of tetrad \cite{Mishra:2025ikg}

\begin{eqnarray}
{\left[\delta e^{a}{}_{\mu}\right]}
=
\left[
\begin{array}{cc}
\Phi & a\,\partial_{\hat{i}}\zeta \\
\delta^{\tilde{l}}{}_{\hat{i}}\partial^{\hat{i}}\zeta &
a\,\delta^{\tilde{l}\hat{i}}\left(\,\Psi\,\delta_{\hat{i}\hat{j}}+\epsilon_{\hat{i}\hat{j}\hat{k}}\partial^{\hat{k}}\sigma\right)
\end{array}
\right].
\label{perttet}
\end{eqnarray}
Where the temporal part of the gauge transformation vector $\xi^\mu$ has been fixed by equating $a^{-1}e^{0}_{~\hat{i}}$ and $\delta_{\hat{i}\hat{j}}\delta_{a}^{~\hat{j}}e^{a}_{~0}$ and the scalar perturbation $E$
that appears in the spatial part of the perturbed co-tetrad as $\partial_{\hat{i}}\partial_{\hat{j}}E$ has been gauged away, which fixes the spatial part of the $\xi^\mu$. Since NGR is parity preserving, the pseudoscalar mode $\sigma$ can be treated separately. But we include this and see its implications in type 3. Similarly, the energy-momentum tensor is perturbed as 
\begin{align}
\delta T_{\mu\nu}
=
(\delta\rho+\delta p)\mathbb{n}_{\mu}\mathbb{n}_{\nu}
+(\rho+p)\left(u_{\mu}\delta u_{\nu}+u_{\nu}\delta u_{\mu}\right)\nonumber\\
+\delta p\,g_{\mu\nu}
+p\,\delta g_{\mu\nu}-\rho_{\Lambda}\delta g_{\mu\nu}\,.
\end{align}
The normalization condition $(\mathbb{n}^\mu+\delta u^{\mu})(\mathbb{n}_\mu+\delta u_{\mu})=-1-2\Phi$ fixes the time component of the fluid velocity as $\delta u^0=-\Phi$ and the scalar part of the spatial section is given by $\delta u^{\hat{i}}=a^{-1}\partial^{\hat{i}}U$. The on-shell linearized equations \eqref{symmetric split} are given by 
\begin{align}
&{\mathbb{E}_{00}} \equiv 3H\tilde{c}_1\left(\dot{\Psi}-H\Phi\right)
+\frac{\tilde{c}_2}{a}\triangle\dot{\zeta}
-\frac{1}{a^2}\left(\tilde{c}_2\triangle\Phi+2c_3\triangle\Psi\right)
=
\kappa^2\delta\rho,\label{ttcomponent}
\\[1ex]
&\int_{\hat{i}}{\mathbb{E}_{0\hat{i}}} \equiv \frac{\tilde{c}_2}{a}\triangle\zeta
+\tilde{c}_1\left(H\Phi-\dot{\Psi}\right)
=
-a\kappa^2U(p+\rho),
\\[1ex]
&\int_{\hat{i}}{\mathbb{E}_{\hat{i}0}} \equiv 3aH\tilde{c}_2\dot{\zeta}
+a\tilde{c}_2\ddot{\zeta}
+2H\tilde{c}_2\Psi
-\tilde{c}_2\dot{\Phi}
+\left(-2c_1+c_2+c_3\right)H\Phi
-2c_3\dot{\Psi}
=
-a\kappa^2U(p+\rho),
\\[1ex]
&\underset{{\hat{i}\neq\hat{j}}}{\mathbb{E}_{{\hat{i}\hat{j}}}}\equiv(2c_1+c_2)\,a^2\,\epsilon_{\hat{i}\hat{j}\hat{k}}\,
\square\!\left(\partial_{\hat{k}}\sigma\right)
+
\partial_{\hat{i}}\partial_{\hat{j}}
\left[
c_3\Phi
+\tilde{c}_3\Psi
-a\,\tilde{c}_2
\left(3H\zeta+\dot{\zeta}\right)
\right] =0\label{anisotropic}
\\[1ex]
& {\mathbb{E}^{\hat{i}}\,_{\hat{i}}} \equiv \tilde{c}_1
\left[
3H^2\Phi
+H(\dot{\Phi}-3\dot{\Psi})
-\ddot{\Psi}
+2\dot{H}\Phi
\right]
+\frac{\tilde{c}_2}{3a}\triangle\dot{\zeta}
+\frac{2}{3a^2}
\left(\tilde{c}_3\triangle\Psi+c_3\triangle\Phi\right)
=
\kappa^2\delta p.
\end{align}
where 
\begin{equation}
\tilde{c}_1=2c_1-c_2+3c_3,\qquad
\tilde{c}_2=2c_1-c_2+c_3,\qquad
\tilde{c}_3=2c_1-c_2+2c_3\,,
\end{equation}
which, in the TEGR limit \eqref{tegrchoice} reduce to $\tilde{c}_1=2,\,
\tilde{c}_2=0,\,
\tilde{c}_3=1.$ We see that in type 3, characterized by \eqref{Type 3 parameter condition}, the first term in the anisotropic part of the equation \eqref{anisotropic} vanishes, dropping the pseudoscalar from the equations. The $\Box=\bar\nabla_\mu\bar\nabla^\mu$ denotes the 4D d'Alembert operator of the Levi-Civita connection and $\triangle=\partial^{\hat{i}}\partial_{\hat{i}} \equiv-k^2$ is the spatial Euclidean Laplacian. This leaves only one non-vanishing component of the antisymmetric part of the equations \eqref{antisymmetric} given by 
\begin{align}
\int_{\hat{i}}{\mathbb{E}_{[\hat{i}0]}} &\equiv \tilde{c}_2\left[H\left(3a\dot{\zeta}-2\Phi+2\Psi\right)
+a\ddot{\zeta}
-\dot{\Phi}
+\dot{\Psi}
-\frac{1}{a}\triangle\zeta\right]\nonumber\\
&=0
\,,\label{antcomponent}
\end{align}
which vanishes in GR as $\tilde{c}_2=0$ in TEGR limit. The system of equations obtained above is not complete, the perturbations of energy-momentum are subject to the conservation condition, whose components are given by
\begin{align}
\delta \,\bar{\nabla}_{\nu}\Theta_{0}\,^{\nu}:\quad&\dot{\delta \rho} + 3 H (\delta p + \delta \rho)+ (p + \rho) \frac{1}{a}\triangle U + 3 \dot{\Psi} (p + \rho)=0\,,\\[1ex]
\delta \,\bar{\nabla}_{\nu}\Theta_{\hat{i}}\,^{\nu}:\quad
&\partial_{\hat{i}}\Big[\delta p+ a (\dot{p} + \dot{\rho}) U +(p + \rho)\bigl(4 a H U + a \dot{U} + \Phi \bigr)\Big]=0\,.
\end{align}
which completes the system of equations.
\section{Evolution of Density Perturbations}\label{sec:pertmatter}
We are interested in the evolution of the overdensity of CDM for which
\begin{equation}
p=0,\qquad \delta p=0\,,
\end{equation}
where we conduct the analysis from the onset of the matter-dominated epoch up to the present cosmological time. This is accomplished by deriving the evolution equations for matter density perturbations at sub-horizon scales and subsequently comparing the resulting predictions with the background-hierarchy constraints in type 3 NGR \cite{Tomonari:2026lir}. The corresponding set of independent equations in Fourier space is given by
\begin{subequations}
\begin{align}
&\frac{k^2}{a^2} \big(c_3 (\Phi + 2 \Psi) - 2 c_2 \Phi \big)
+ \left(2 c_2 - c_3\right)\frac{k^2}{a} \dot{\zeta}
+ 3 \left(2 c_2 - 3 c_3\right) H \big(H \Phi - \dot{\Psi}\big)
= \kappa^2 \, \delta \rho \,, \label{perttt}\\[6pt]
&\left(2 c_2 - c_3\right)\frac{k^2}{a} \zeta
+ \left(3 c_3 - 2 c_2\right) \left(H \Phi - \dot{\Psi}\right)
=  \kappa^2 \rho V \,,\label{paniso} \\[6pt]
&3 a H \dot{\zeta} + \frac{ \, k^2}{a}\zeta + a \ddot{\zeta}
- 2 H \Phi + 2 H \Psi - \dot{\Phi} + \dot{\Psi} = 0 \,,\label{pantieq} \\[6pt]
&a \left(c_3 - 2 c_2\right) \left( \dot{\zeta}+3 H \zeta \right)
- c_3 \Phi + 2 \left(c_2 - c_3\right) \Psi = 0 \,, \label{pertij}\\[6pt]
& \left(2 c_2 - c_3\right) \frac{k^2}{a} \left( \dot{\zeta}+2 H\zeta \right)
-  \left(2 c_2 - 3 c_3\right) \left(3 H^2 \Phi + H \dot{\Phi} - 3 H \dot{\Psi}
+ 2 \dot{H} \Phi - \ddot{\Psi}\right) = 0 \,,\label{pertii}\\
&\dot{\delta \rho} + 3 H  \delta \rho +\frac{k^2}{a^2}\rho V + 3 \rho\dot{\Psi}=0\,,\\
&\dot{V}-\Phi=0\,.\label{veq}
\end{align}
\end{subequations}
where $V=-aU$. The equation \eqref{pertii} is obtained by removing the traceless part from the trace equation. To obtain the last two equations, the background matter conservation equation \eqref{matterconservationbackground} has been used. The antisymmetric part \eqref{pantieq} has been obtained by considering $\tilde{c}_2\neq0$, which in type 3 NGR reduces to $c_3-2c_2\neq0$. Therefore, all the coefficients of $\zeta$ and its derivatives survive. The last two equations in terms of gauge invariant matter density perturbation $\delta_m=\displaystyle\frac{{\delta\rho}}{\rho}+3HV$ can be combined as
\begin{equation}
\dot{\delta}_m +V \left(\frac{k^2}{a^2}-3 \dot{H}\right)-3 H\Phi+3 \dot{\Psi}=0
\end{equation}
using which we derive the evolution equation
\begin{equation}
\begin{aligned}
\ddot{\delta}_m+2 H \dot{\delta}_m
+\Phi \left(\frac{k^2}{a^2}-6 H^2-6 \dot{H}\right)-3V (2 H \dot{H}+\ddot{H})
-3 H (\dot{\Phi}-2 \dot{\Psi})+3 \ddot{\Psi}=0\,.
\end{aligned}
\label{matterevolution}
\end{equation}

\subsection{Subhorizon Modes}
Within the sub-horizon approximation (SHA), the relevant modes possess physical momenta $k/a$ that are much larger than the Hubble expansion rate $H$. For modes well inside the horizon, we additionally invoke a quasi-static approximation (QSA), which consists of neglecting the time derivatives of the cotetrad perturbations. We adopt the approach of \cite{Mishra:2026pjz}, and define the SHA and QSA parameters respectively as
\begin{equation}
\label{sha}
\varepsilon \equiv \frac{a H}{k}, \quad 
\delta \equiv \frac{\dot{\varepsilon}}{\varepsilon H}, \quad 
\xi \equiv \frac{\ddot{\varepsilon}}{\varepsilon H^2}, \quad  
\chi \equiv \frac{\dddot{\varepsilon}}{\varepsilon H^3},
\end{equation}
and
\begin{subequations}
\label{qsa}
\begin{align}
\varepsilon_\Phi \equiv \frac{\dot{\Phi}}{H\Phi }, \qquad 
\varepsilon_\Psi \equiv \frac{\dot{\Psi}}{H\Psi }, \qquad 
\varepsilon_\zeta \equiv \frac{\dot{\zeta}}{H\zeta },\\
\chi_\Phi \equiv \frac{\dot{\varepsilon}_\Phi}{H\varepsilon_\Phi }, \qquad 
\chi_\Psi \equiv \frac{\dot{\varepsilon}_\Psi}{H\varepsilon_\Psi },\qquad
\chi_\zeta \equiv \frac{\dot{\varepsilon}_\zeta}{H\varepsilon_\zeta }.
\end{align}
\end{subequations}
Accordingly, the sub-Hubble condition $k\gg aH$ corresponds to $\varepsilon\ll1$, while $\dot{\Phi}\approx0$ implies $\varepsilon_\Phi\ll1$. Unlike the QSA parameters, the SHA parameters other than $\varepsilon$ need not be small. For example, during matter domination, $H\propto a^{-3/2}$, giving $\displaystyle\delta=-\frac{1}{2},\xi=1,\, \text{and}\, \chi=-\frac{7}{2}$. To implement this parametrization, the perturbed system of equations is expanded in the perturbation variables and their time derivatives, retaining only the leading terms in $\varepsilon$. Since the QSA parameters are small, their products with $\varepsilon$ are higher-order contributions and can be neglected. Likewise, higher derivatives of $\varepsilon_\Phi$ are nonlinear in the QSA parameters and are omitted. In what follows, the equalities are up to order $\varepsilon^2$. Expanding the antisymmetric part in these parameters and solving for the Lorentz perturbation $\zeta$, we get 
\begin{equation}
\zeta =\frac{\varepsilon  \left(\Phi  \left(\varepsilon _{\Phi }+2\right)-\Psi  \left(\varepsilon _{\Psi }+2\right)\right)}{k(\varepsilon ^2 \varepsilon _{\zeta } \left(\delta +\varepsilon _{\zeta }+\chi _{\zeta }+2\right)+1)}\,.\label{zetarm}
\end{equation}
Substituting this into \eqref{pertij}, we get at first order
\begin{equation}
\left(2 \left(c_2-c_3\right) \Psi -c_3 \Phi \right)\simeq0\,.\label{pijeq}
\end{equation}
Next, we multiply the time–space component by $3H$ and add the result to the time–time component, and subsequently substitute the expression for $\zeta$ from Eq.~\eqref{zetarm}. At leading order, this yields
\begin{equation}
\left(c_3-2 c_2\right) \frac{k^2}{a^2} \Phi +2 c_3 \frac{k^2}{a^2} \Psi -\kappa ^2 \rho  \delta _m\simeq0\,.\label{pttqsa}
\end{equation}
The evolution equation \eqref{matterevolution} under this approximation becomes
\begin{equation}
\ddot{\delta}_m + \frac{2 k \varepsilon \dot{\delta}_m}{a} + \frac{k^2 \Phi}{a^2} \simeq 0.
\end{equation}
which after substituting $\varepsilon=\frac{a H}{k}$ is 
\begin{equation}
\ddot{\delta}_m + 2 H\dot{\delta}_m + \frac{k^2 \Phi}{a^2} \simeq 0\,.\label{matterqsa}
\end{equation}
Solving simultaneously \eqref{pijeq}, \eqref{pttqsa} and \eqref{matterqsa}, we get
\begin{equation}
\ddot{\delta}_m+2 H \dot{\delta}_m+ \frac{\left(c_3-c_2\right) \kappa ^2 \rho  }{c_2 \left(2 c_2-3 c_3\right)}\delta _m=0\,.\label{meszorasngr}
\end{equation}
The above equation is the M\'esz\'aros  equivalent of type 3 NGR with a cosmological constant. This reduces to the standard $\Lambda$CDM result for TEGR limit \eqref{tegrchoice}.  Comparing this with the standard matter perturbation equation 
\begin{equation}
\ddot{\delta}_m+2 H \dot{\delta}_m- \frac{\kappa^2}{2}\rho\frac{G_{\rm eff}}{G}\delta _m=0\,.\label{meszorasstandard}
\end{equation}
we get 
\begin{equation}
G_{\mathrm{eff}}=\frac{2 \left(c_2-c_3\right) G}{c_2 \left(2 c_2-3 c_3\right)}\,. \label{geff}
\end{equation}
Reduction of the $f\sigma_8$ (or $S_8$) tension in modified-gravity scenarios generically requires a late-time suppression of the effective gravitational constant, $G_{\rm eff}<G$, so that structure growth is weakened without strongly altering the background expansion \cite{Nesseris:2017vor,Kazantzidis:2018rnb,Anagnostopoulos:2026fQ,Perivolaropoulos:2025growth,Bernal:2026MG}. Recall that the no-instability and background hierarchy bounds in the perturbative sector of type 3 NGR are given by 
\begin{equation}
c_2>0,\qquad 2c_2<c_3<4c_2\,.\label{bghrb}
\end{equation}
Fixing the background \eqref{background} to reproduce the standard Friedmann equations leads to the constraint $2c_2 - 3c_3 = -2$. Under this condition, the expression \eqref{geff} simplifies to
\begin{equation}
G_{\mathrm{eff}}=\frac{ \left(c_3-c_2\right) }{c_2}G\,. \label{geffc2}
\end{equation}
Thus $\displaystyle\frac{G_{\mathrm{eff}}}{G}<1$ implies that $c_3<2c_2$, which with the requirement \eqref{bghrb} is impossible. Note that fixing the background expansion to be that of $\Lambda$CDM leaves one of the coefficients $c_2$ and $c_3$ arbitrary. Therefore, type 3 NGR is not able to reduce $f\sigma_8$ (or $S_8$) tension. This is the central result of the present study. In the next section, we use the growth rate data to assess dataset preferences based on constraints derived from the model parameters. Before doing so, we also specify the other two parameters that specify deviation from GR, known as gravitational slip and the lensing-normalized Newton constant given by \cite{CANTATA:2021ktz}
\begin{align}
\Psi =-\eta(a,k) \Phi\,,\label{gslip}\\
k^2 (\Phi -\Psi )=- a^2 \kappa ^2 \rho  \delta _m \Sigma (a,k)\,. \label{lensing}
\end{align}
From \eqref{pijeq}, we get 
\begin{equation}
\eta(a,k)=\frac{1+c_2}{2-c_2}\,,
\end{equation}
and eliminating $\zeta$ and $\dot{\zeta}$ from \eqref{perttt}, \eqref{paniso} and \eqref{pertij}, we get 
\begin{equation}
\Sigma(a,k)=\frac{1}{2 c_2}\,,
\end{equation}
which satisfy the standard relation 
\begin{equation}
\Sigma(a,k)=\frac{G_{\rm eff}(1+\eta(a,k))}{2G}\,,
\end{equation}
while in GR, they are equal to unity. We see that in NGR, all these parameters are constants and depend on the model parameter $c_2$. In the above calculations, we have used $2c_2 - 3c_3 = -2$ to express the equations in terms of the free parameter $c_2$. The equation \eqref{meszorasstandard}, is simplified to 
\begin{equation}
\ddot{\delta} _m+2 H \dot{\delta }_m-\frac{\kappa ^2\left(2-c_2\right)  \rho  \delta _m}{6c_2}=0\,.\label{modimeszorasc2}
\end{equation}
Before further analysis, we evaluate the phenomenological parameters across the theoretically viable parameter space.
With $2c_2-3c_3=-2$ one has $c_3=(2c_2+2)/3$, so that $c_3>2c_2$ (ghost-free) is equivalent to $c_2<1/2$ and
$c_3<4c_2$ (hierarchy) to $c_2>1/5$.
In the viable range $0.2<c_2<0.5$ the parameters take the values
\begin{subequations}
\begin{align}
\mu\equiv\frac{G_{\rm eff}}{G}
&=\frac{2-c_2}{3c_2}\in(1,3)\,,\\
\eta
&=\frac{1+c_2}{2-c_2}\in\left(\frac23,1\right),\\
\Sigma
&=\frac{1}{2c_2}\in\left(1,\frac52\right).
\end{align}
\end{subequations}
with $\Sigma-1=\tfrac34(\mu-1)$.
Redshift-space distortions are sensitive to the growth parameter $\mu$, while weak lensing is sensitive to $\Sigma$ as well as to the growth history. Because both exceed unity in the viable range, the growth rate and the lensing signal are simultaneously enhanced. This raises both the $f\sigma_8$ and the $S_8$ predictions above those of $\Lambda$CDM, whereas the data prefer lower values. Suppressing either observable requires $c_2>1/2$, which is precisely the region excluded by the ghost-free condition.

\subsection{Growth Factor}
\begin{figure}[H]
\centering
\includegraphics[width=\linewidth]{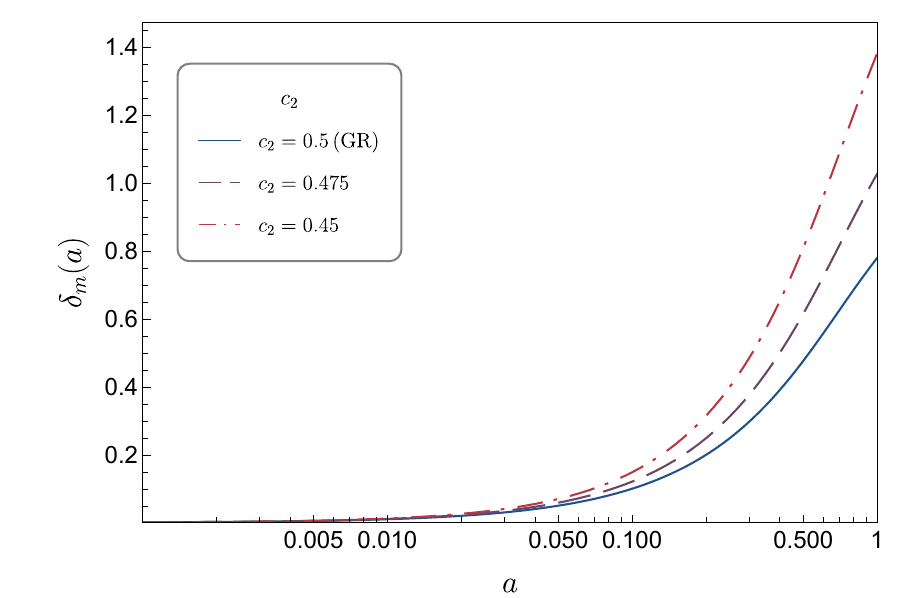}
\caption{Evolution of $\delta_m$ for $\Omega_{m0}=0.3$.}
\label{fig:growth}
\end{figure}

Changing the independent variable to be the scalar factor and using $\displaystyle\rho=\frac{3H_0^2}{\kappa^2}\Omega_{m0}a^{-3}$, \eqref{modimeszorasc2} is transformed to
\begin{equation}
\frac{d^2 \delta _m}{da^2}+\left(\frac{3}{a}+\frac{1}{H}\frac{d H}{da}\right) \frac{d \delta _m}{da}- \frac{\left(2-c_2\right) \Omega _{m0}H_0^2}{2 a^5 H^2 c_2} \delta _m=0\,.
\end{equation}
Which transforms to 
\begin{equation}\label{mesroasina}
\frac{d^2 \delta _m}{da^2}+3\left(\frac{1}{a}-\frac{1}{2a \left(1+a^3 \mathcal{R}_m\right)}\right) \frac{d \delta _m}{da}- \frac{\left(2-c_2\right) \delta _m}{2 a^2 c_2\left(1+a^3 \mathcal{R}_m\right)}=0\,.
\end{equation}
under the background equation $\displaystyle H^2=H_0^{2}\left(\Omega_{m0} a^{-3}+\Omega_{\Lambda}\right)=H_0^{2}\left(\Omega_{m0} a^{-3}+\mathcal{R}_m\Omega_{m0}\right)$, and $\mathcal{R}_m=\frac{\Omega_{\Lambda}}{\Omega_{m0}}=\frac{(1-\Omega_{m0})}{\Omega_{m0}}$. 
The growing-mode solution of Eq.~\eqref{mesroasina} can be expressed
in closed form in terms of the Gauss hypergeometric function. We give
the explicit derivation in Appendix~\ref{app:growth} and quote only
the result here:
\begin{equation}
  \delta_m(a;c_2) = \mathcal{A}\, a^{\alpha}\,
  {}_2F_1\!\left(\frac{\alpha}{3},\, \frac{\alpha+2}{3};\,
  \frac{4\alpha+7}{6};\, -a^3 \mathcal{R}_m\right),
  \label{eq:growth}
\end{equation}
with
\begin{equation}
  \alpha = \frac{-1 + \sqrt{-7 + 16/c_2}}{4}.
\end{equation}
For $c_2 = 1/2$, $\alpha = 1$ and this reduces to the standard
$\Lambda$CDM growth factor
$D_+(a) = a\,{}_2F_1(1/3, 1; 11/6; -a^3\mathcal{R}_m)$. Fixing the integration constant $\mathcal{A}$ by using the initial condition $\delta_m(a_{ini})=a_{ini}$ at $a_{ini}=0.001$, we show the evolution of $\delta_m$ with respect to the scale factor in Fig.~\ref{fig:growth}. It is clear that admissible values of $c_2$ cannot reduce the growth of matter overdensity at late times.
\section{Observational consistency with \texorpdfstring{$f\sigma_8$}{fσ8} data}
\label{sec:fs8}

As a consistency check of the analytic results of Sec.~\ref{sec:pertmatter},
we confront the growth history predicted by Type-3 NGR with measurements
of the redshift-space growth rate
\begin{equation}
  f\sigma_8(a) \equiv a\,\sigma_8\,\frac{1}{\delta_m(a=1)}
  \frac{d\delta_m}{da},
  \label{fs8}
\end{equation}
where $\sigma_8$ is the present-day root mean square matter fluctuation over $8\,h^{-1}\,{\rm Mpc}$. We use the $22$ measurements compiled in Table~\ref{tab:fs8_measurements} (Appendix \ref{app:data}) from Refs.~\cite{Kazantzidis:2018rnb,Sagredo:2018ahx,Benisty:2020kdt,Hudson:2012gt,Huterer:2016uyq,Feix:2015dla,Howlett:2014opa,Song:2008qt,Blake:2013nif,Samushia:2011cs,BOSS:2013eso,2012MNRAS.425..405B,BOSS:2013mwe,Pezzotta:2016gbo,eBOSS:2018yfg,Okumura:2015lvp}.
The measurements from WiggleZ~\cite{2012MNRAS.425..405B} and SDSS-IV
eBOSS~\cite{eBOSS:2018yfg} are treated with their full published
covariance matrices, while all other data points are taken as
independent.

\begin{figure}[htbp]
  \centering
  \includegraphics[width=0.9\linewidth]{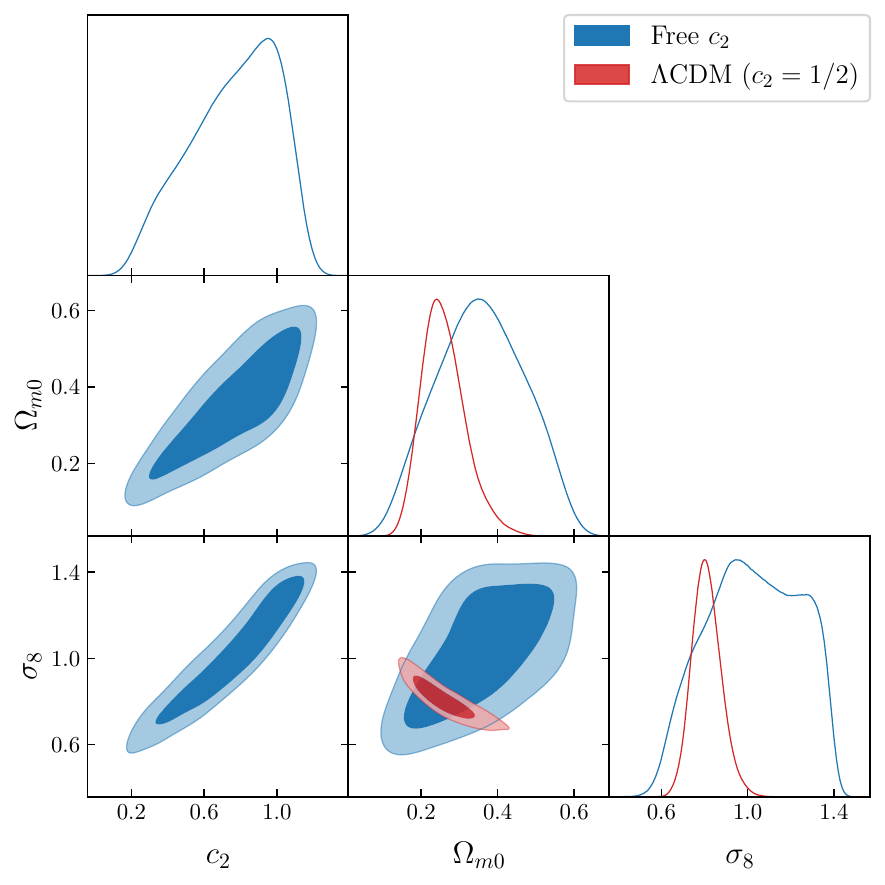}
  \caption{Posterior distributions of the free-$c_2$ model (blue) and 
    the $\Lambda$CDM reference with $c_2 = 1/2$ fixed (red). The 
    free-$c_2$ model yields a best fit at $c_2 = 0.57$ and a median of 
    $c_2 = 0.76 \pm 0.24$, consistent with the $\Lambda$CDM value at 
    the $1\sigma$ level. The $\sigma_8$ marginal broadens by a factor 
    of three and extends toward the prior bound at $\sigma_8 = 1.4$.}
  \label{fig:corner}
\end{figure}

\begin{figure}[htbp]
  \centering
  \includegraphics[width=0.75\linewidth]{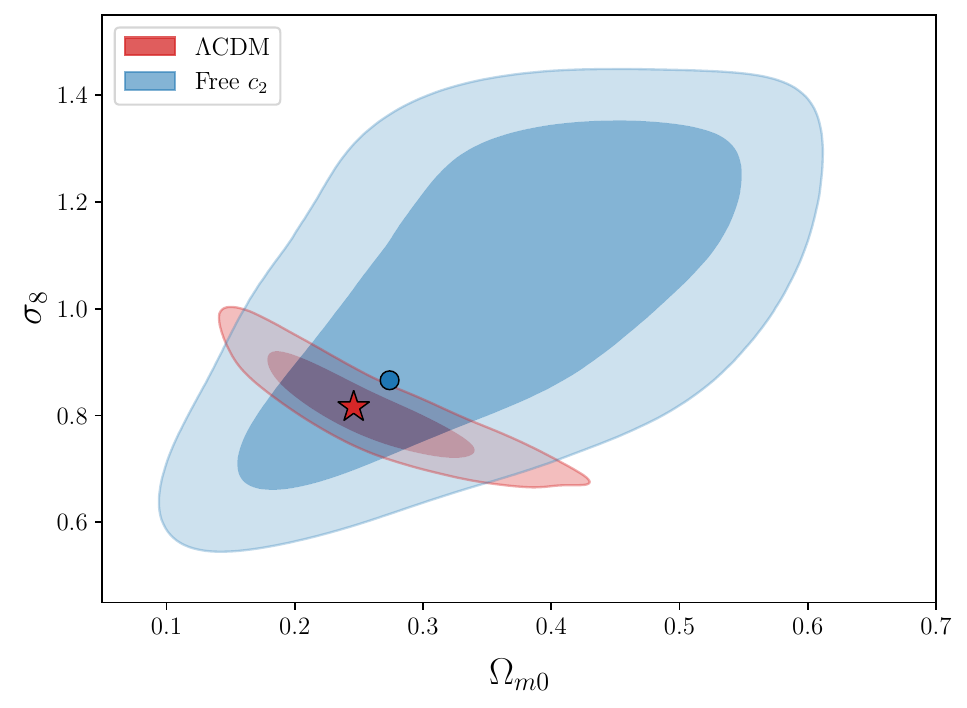}
  \caption{Constraints on $(\Omega_{\rm m0},\sigma_8)$ from the $f\sigma_8$ data. Filled contours show the $68\%$ and $95\%$ credible regions for the free-$c_2$ model (blue) and $\Lambda$CDM (red). The star and circle mark their respective best-fit points. The two constraints have opposite orientations: $\Lambda$CDM follows the standard negative degeneracy, while the free-$c_2$ model follows a new positive-degeneracy direction. The broadening of the blue contour reflects a parameter degeneracy rather than a preference for shifted values, with the $\Lambda$CDM solution sitting at the $1\sigma$ credible level of the free-$c_2$ posterior.}
  \label{fig:overlay}
\end{figure}

We sample the parameter space with the affine-invariant ensemble sampler \texttt{emcee}~\cite{ForemanMackey:2012ig}, using the likelihood $\mathcal{L}\propto e^{-\chi^2/2}$ with correlations included through the covariance matrices of the correlated data. Uniform priors are adopted on $\Omega_{\rm m0}\in[0.1,0.6]$ and $\sigma_8\in[0.4,1.4]$. The parameter $c_3$ is fixed by the background
condition $2c_2 - 3c_3 = -2$ of Sec.~\ref{sec:pertmatter}, so that the
background expansion matches that of $\Lambda$CDM for all $c_2$. We
run two chains. The reference $\Lambda$CDM chain has $c_2 = 1/2$
fixed, and the free-$c_2$ chain adopts the uniform prior $0<c_2<2$.
The upper bound at $c_2 = 2$ is the value at which
$G_{\mathrm{eff}} = 0$, beyond this point the effective gravitational
coupling changes sign and the growth equation is no longer physical.
The prior deliberately includes both the theory-viable region \(0.2 < c_2 < 0.5\) and the tension-reducing region \(0.5 < c_2 < 2\), to capture the full likelihood shape across both regimes. Convergence is confirmed via the
integrated autocorrelation time, with $N/\tau>50$ and stable $\tau$
reached at step $1900$ for $\Lambda$CDM ($\tau\simeq 32$) and at
step $3500$ for free $c_2$ ($\tau\simeq 62$). The marginalized $68\%$ credible intervals and best-fit parameters
are summarized in Table~\ref{tab:mcmc}. The free-$c_2$ posterior is
shown in Fig.~\ref{fig:corner}, and the projection onto the
$(\Omega_{\rm m0},\sigma_8)$ plane with the $\Lambda$CDM reference is
shown in Fig.~\ref{fig:overlay}.
\begin{table}[htbp]
\centering
\caption{\label{tab:mcmc}
Constraints from the $f\sigma_8$ data. Best-fit parameters, minimum
$\chi^2$, degrees of freedom, marginalized $68\%$ credible intervals,
and log-evidence. The third column is not a measurement of
$c_2$ but a shape diagnostic with $\Omega_{\rm m0}$, $\sigma_8$
and $c_2$ all free.}
\begin{tabular}{lcc}
\hline\hline
 & $\Lambda$CDM ($c_2=\tfrac12$) & Free $c_2$ (Diagnostic) \\[0.1cm]
\hline
Best fit: $c_2$                  & $0.5$              & $0.57$ \\
Best fit: $\Omega_{\rm m0}$      & $0.246$            & $0.274$ \\
Best fit: $\sigma_8$             & $0.816$            & $0.866$ \\
$68\%$ CL: $\Omega_{\rm m0}$     & $0.26 \pm 0.06$    & $0.36 \pm 0.12$ \\
$68\%$ CL: $\sigma_8$            & $0.81 \pm 0.07$    & $1.03 \pm 0.21$ \\
$68\%$ CL: $c_2$                 & ---                & $0.76 \pm 0.24$ \\
$\chi^2_{\rm min}$               & $12.364$           & $12.354$ \\
d.o.f.                           & $20$               & $19$ \\
$\chi^2_{\rm min}/\mathrm{d.o.f.}$ & $0.618$          & $0.650$ \\
$\ln Z$                          & $-10.06$           & $-10.53$ \\
\hline\hline
\end{tabular}
\end{table}

Three features of these results are relevant for the interpretation.

\paragraph{Unconstrained shape.}The $f\sigma_8$ shape alone does not constrain $c_2$.
With $\Omega_{\rm m0}$ and $\sigma_8$ both free, the profile likelihood over $c_2$ is flat: $\Delta\chi^2 < 1$ for $0.2 < c_2 < 0.8$. It rises only where the fit reaches the prior edges $\Omega_{\rm m0} = 0.6$ or $\sigma_8 = 1.4$. Those bounds are not supported by external data (CMB and weak-lensing analyses give $\sigma_8 \approx 0.75$--$0.85$), and they set the upper edge of the $c_2$ posterior. Neither the median nor the location of the minimum is therefore a measurement of $c_2$.

The Bayes factor confirms this interpretation. Table~\ref{tab:bayes_prior} shows the Savage-Dickey Bayes factor $\ln B(\Lambda{\rm CDM}:c_2)$ computed from the same chain for three different prior ranges. It changes from positive (favoring $\Lambda$CDM by a factor $1.6$) to negative (favoring the free-$c_2$ model by a factor $1.4$) as the prior narrows. The factor is not a robust data preference but an Occam penalty set by the prior volume. We draw no model preference from it.

\begin{table}[htbp]
\centering
\caption{\label{tab:bayes_prior}
Savage-Dickey Bayes factor $\ln B(\Lambda{\rm CDM}:c_2)$ for three
different prior ranges on $c_2$. The factor changes from positive to
negative as the prior narrows, confirming that it reflects prior
volume rather than a data preference.}
\begin{tabular}{lcc}
\hline\hline
$c_2$ prior range & $\ln B(\Lambda{\rm CDM}:c_2)$ & $B$ \\
\hline
$(0, 2)$     & $+0.45$ & $1.56$ \\
$(0, 1.2)$   & $-0.06$ & $0.94$ \\
$(0.2, 0.5)$ & $-0.33$ & $0.72$ \\
\hline\hline
\end{tabular}
\end{table}
\paragraph{Degeneracy rather than measurement.} The likelihood maximum lies
at $c_2 = 0.57$, only $0.07$ above the $\Lambda$CDM value of $1/2$,
with an improvement of $\Delta\chi^2 = 0.011$ ($0.1\sigma$). Meanwhile
the marginalized $\Omega_{\rm m0}$ and $\sigma_8$ shift upward while
their posterior widths inflate substantially. This is visible in
Table~\ref{tab:mcmc} as $\sigma_8 = 0.81 \pm 0.07$ for $\Lambda$CDM
versus $\sigma_8 = 1.03 \pm 0.21$ for free $c_2$, a factor of three
inflation, and in Fig.~\ref{fig:overlay}, where the $\Lambda$CDM point
sits at the $1\sigma$ credible level of the free-$c_2$ posterior. The
free-$c_2$ posterior median sits at $c_2 \simeq 0.76$, above the
likelihood maximum $c_2 = 0.57$, because the posterior is dragged
upward by the $\sigma_8$ prior bound discussed below. The shifts in
the means are at the $\sim 1\sigma$ level and reflect this parameter
degeneracy rather than a statistical preference.

The $\sigma_8$ marginal for the free-$c_2$ model extends toward the
upper prior bound at $\sigma_8 = 1.4$, reflecting a genuine degeneracy
rather than a numerical artifact. Increasing $c_2$ suppresses the
growth rate through $G_{\mathrm{eff}}$, so the data can be fit at
higher $\sigma_8$ provided $c_2$ increases correspondingly, and along
this ridge the likelihood is nearly flat. We truncate the prior at
$\sigma_8 = 1.4$, which is already far above the values
consistent with independent cosmological probes allow ($\sigma_8\simeq 0.75-0.85$). We verified that widening this bound further shifts the posterior median along the
ridge without changing the likelihood maximum or $\Delta\chi^2$, and
confirm that the bound-filling is the quantitative signature of an
unconstrained parameter rather than a numerical artifact. We therefore
retain $\sigma_8 < 1.4$ as a reasonable and sufficient prior.

\paragraph{The tension-reducing region is theoretically excluded.}
Reducing the $S_8/f\sigma_8$ tension requires a suppressed growth rate, i.e. $G_{\rm eff} < G$, which holds if and only if $c_2 > 1/2$. However, the same condition $c_2 > 1/2$ is incompatible with the ghost-free and hierarchy bounds~\eqref{bghrb}. Combining $2c_2 < c_3 < 4c_2$ with $2c_2 - 3c_3 = -2$ yields the theory-viable range $0.2 < c_2 < 1/2$, where $G_{\rm eff}/G > 1$ and the growth rate is enhanced rather than suppressed relative to $\Lambda$CDM. The region of parameter space that would relieve the tension is therefore excluded by the same stability requirements that render the theory viable. As established above, the data cannot distinguish the models and provide no evidence for a departure from $\Lambda$CDM. No viable choice of $c_2$ reconciles Type-3 NGR with the $S_8/f\sigma_8$ tension, in agreement with the analytic result of Sec.~\ref{sec:pertmatter}.

\section{Conclusions}
\label{sec:conclusions}

We have examined whether Type-3 New General Relativity can reduce the $S_8/f\sigma_8$ cosmic tension. We derived the scalar perturbation equations in the presence of a cosmological constant, reduced them under sub-horizon and quasi-static approximations to a M\'esz\'aros-type equation, and extracted the effective gravitational coupling on a background fixed to $\Lambda$CDM. We then imposed the background hierarchy bounds on the resulting parameter space and confronted the predicted growth history with twenty-two $f\sigma_8$ measurements from WiggleZ, SDSS-IV eBOSS, BOSS, VIPERS, and other surveys, using an MCMC analysis with both a fixed and a free value of $c_2$.

The analytic result is the decisive one. The parameter region in which the growth rate is suppressed relative to $\Lambda$CDM is exactly the region in which the theory develops instabilities. Viability bounds do not merely narrow the admissible parameter space, but remove the tension-reducing region entirely. Stability and tension reduction are therefore mutually exclusive by construction. The observational analysis corroborates this conclusion without being decisive on its own. With $\Omega_{\rm m0}$ and $\sigma_8$ free, the $f\sigma_8$ data cannot distinguish the models. The Bayes factor changes sign with the prior range, and the posterior is dominated by prior edges. The data offers no rescue for the tension, and no viable choice of $c_2$ can suppress both the growth rate and the lensing signal simultaneously.

Several directions follow. The most direct is to ask whether the same obstruction appears in other NGR subclasses, or whether it is specific to Type-3. A second is to implement the full perturbation equations in a Boltzmann solver, so that joint CMB, BAO, and supernova constraints can be applied without the sub-horizon approximations used here. 
\appendix
\section{Data and Statistical Framework}
\label{app:data}

\subsection{Bayesian Inference}

The observational constraints in this work are derived using Bayesian parameter estimation. Given a dataset $\{x_i\}$ and a model parametrized by $\theta$, the posterior probability distribution $P(\theta|\{x_i\})$ is given by Bayes' theorem:
\begin{equation}
    P(\theta|\{x_i\}) \propto \mathcal{L}(\{x_i\}|\theta)\,\pi(\theta),
\end{equation}
where $\mathcal{L}(\{x_i\}|\theta)$ is the likelihood of the data given the parameters, and $\pi(\theta)$ is the prior probability distribution reflecting our theoretical knowledge or assumptions about $\theta$.

For the $f\sigma_8$ measurements used in this study, the likelihood is assumed to be Gaussian in the residuals:
\begin{equation}
  \mathcal{L}(\theta) \propto
  \exp\!\left[-\frac{1}{2}\chi^{2}(\theta)\right],
\end{equation}
with
\begin{equation}
  \chi^{2}(\theta) =
  \sum_{i,\,j}
  \left[x_{i}^{\rm th}(\theta)-x_{i}^{\rm obs}\right]
  \left([C^{-1}]_{ij}\right)
  \left[x_{j}^{\rm th}(\theta)-x_{j}^{\rm obs}\right],
\end{equation}
where $x_{i}^{\rm obs}$ is the $i$-th measured value of $f\sigma_8$, $x_{i}^{\rm th}(\theta)$ is the corresponding model prediction, and $C_{ij}$ is the covariance matrix. For uncorrelated points, $C_{ij} = \sigma_{i}^{2}\,\delta_{ij}$, reducing the expression to the standard sum over independent residuals. For correlated data, the full covariance matrices are used. The total $\chi^{2}$ is the sum of the independent and correlated components:
\begin{equation}
  \chi^{2} = \chi^{2}_{\rm diag}
           + \chi^{2}_{\rm WiggleZ}
           + \chi^{2}_{\rm SDSS}.
\end{equation}
\subsection{Bayes factor and model comparison}
\label{app:bayes}

Model selection between $\Lambda$CDM (with $c_2 = 1/2$ fixed) and the
free-$c_2$ model is quantified through the Bayes factor
\begin{equation}
  B(\Lambda{\rm CDM} : c_2) = \frac{Z_{\Lambda{\rm CDM}}}{Z_{c_2}},
  \qquad
  Z \equiv \int \mathcal{L}(\theta)\,\pi(\theta)\,d\theta,
  \label{eq:bayesfactor}
\end{equation}
where $Z$ is the marginalized likelihood of each model. Throughout the paper $\ln B$ denotes the natural logarithm of the Bayes factor, with positive values favoring $\Lambda$CDM.

Because the two models are nested and the priors on the shared parameters are identical, $B$ can be evaluated from the free-$c_2$ chain alone using the Savage--Dickey density ratio,
\begin{equation}
  B(\Lambda{\rm CDM} : c_2)
  = \frac{p(c_2 = 1/2 \mid \text{data})}{\pi(c_2 = 1/2)}.
  \label{eq:savage_dickey}
\end{equation}
The posterior density is estimated from the post-burn-in samples with a
Gaussian kernel density estimator, and the bootstrap uncertainty on
$\ln B$ is obtained by resampling the chain. A direct grid integration
of Eq.~\eqref{eq:bayesfactor} agrees with the Savage--Dickey result
within that uncertainty. For a flat prior over
$[c_2^{\min}, c_2^{\max}]$, the denominator of
Eq.~\eqref{eq:savage_dickey} scales as $1/(c_2^{\max} - c_2^{\min})$.
The change of sign of $\ln B$ in Table~\ref{tab:bayes_prior} as the
prior narrows is dominated by this prior-volume factor rather than by
a shift in the data-driven posterior.
\subsection{Data Compilation}
The observational constraints are derived from a compilation of
$22$ $f\sigma_8(z)$ measurements spanning $0.02 \leq z \leq 1.944$, listed
in Table~\ref{tab:fs8_measurements}. The data points are taken from
the references shown in the table. All measurements are treated as 
independent except for the WiggleZ and SDSS-IV eBOSS subsets, which 
are correlated and utilize the full covariance matrices provided below.
\begin{table}[htbp]
\centering
\small
\caption{\label{tab:fs8_measurements}
$f\sigma_8$ measurements from various surveys.}
\begin{tabular}{cllc}
\hline\hline
$z_{\rm eff}$ & $f\sigma_8(z)$ & Survey & Reference \\
\hline
0.02  & $0.398 \pm 0.065$ & SN Ia IRAS & \cite{Hudson:2012gt} \\
0.02  & $0.314 \pm 0.048$ & 2MRS & \cite{Hudson:2012gt} \\
0.02  & $0.428 \pm 0.0465$ & 6dFGS+SN Ia & \cite{Huterer:2016uyq} \\
0.10  & $0.37 \pm 0.13$ & SDSS-veloc & \cite{Feix:2015dla} \\
0.15  & $0.490 \pm 0.145$ & SDSS-MGS & \cite{Howlett:2014opa} \\
0.17  & $0.51 \pm 0.06$ & 2dFGRS & \cite{Song:2008qt} \\
0.18  & $0.36 \pm 0.09$ & GAMA & \cite{Blake:2013nif} \\
0.25  & $0.3512 \pm 0.0583$ & SDSS-LRG-200 & \cite{Samushia:2011cs} \\
0.32  & $0.384 \pm 0.095$ & BOSS LOWZ & \cite{BOSS:2013eso} \\
0.37  & $0.4602 \pm 0.0378$ & SDSS-LRG-200 & \cite{Samushia:2011cs} \\
0.38  & $0.44 \pm 0.06$ & GAMA & \cite{Blake:2013nif} \\
0.44  & $0.413 \pm 0.08$ & WiggleZ & \cite{2012MNRAS.425..405B} \\
0.59  & $0.488 \pm 0.06$ & SDSS-CMASS & \cite{BOSS:2013mwe} \\
0.60  & $0.39 \pm 0.063$ & WiggleZ & \cite{2012MNRAS.425..405B} \\
0.60  & $0.550 \pm 0.120$ & VIPERS PDR-2 & \cite{Pezzotta:2016gbo} \\
0.73  & $0.437 \pm 0.072$ & WiggleZ & \cite{2012MNRAS.425..405B} \\
0.86  & $0.40 \pm 0.11$ & VIPERS PDR-2 & \cite{Pezzotta:2016gbo} \\
0.978 & $0.379 \pm 0.176$ & SDSS-IV eBOSS & \cite{eBOSS:2018yfg} \\
1.230 & $0.3850 \pm 0.0990$ & SDSS-IV eBOSS & \cite{eBOSS:2018yfg} \\
1.40  & $0.482 \pm 0.116$ & FastSound & \cite{Okumura:2015lvp} \\
1.526 & $0.342 \pm 0.070$ & SDSS-IV eBOSS & \cite{eBOSS:2018yfg} \\
1.944 & $0.364 \pm 0.106$ & SDSS-IV eBOSS & \cite{eBOSS:2018yfg} \\
\hline\hline
\end{tabular}
\end{table}
\subsection{Covariance Matrices}
Two subsets of the compilation are statistically correlated. For the three WiggleZ data points at $z = 0.44$, $0.60$, and $0.73$, the covariance matrix is~\cite{2012MNRAS.425..405B,Kazantzidis:2018rnb}
\begin{equation}
  \label{wigglecov}
  \underset{{\rm WiggleZ}}{\left[C_{ij}\right]} = \begin{pmatrix}
    0.00640 & 0.002570 & 0.000000 \\
    0.00257 & 0.003969 & 0.002540 \\
    0.00000 & 0.002540 & 0.005184
  \end{pmatrix}.
\end{equation}
For the four SDSS-IV eBOSS data points at $z = 0.978$, $1.230$, $1.526$, and $1.944$, the covariance matrix is~\cite{eBOSS:2018yfg,Sagredo:2018ahx,daSilva:2020mvk}
\begin{align}
  \label{sdsscov}
  \underset{{\rm SDSS}}{\left[C_{ij}\right]} = \begin{pmatrix}
    0.0310 & 0.0089 & 0.0033 & -0.0002 \\
    0.0089 & 0.0098 & 0.0044 &  0.0008 \\
    0.0033 & 0.0044 & 0.0049 &  0.0035 \\
   -0.0002 & 0.0008 & 0.0035 &  0.0112
  \end{pmatrix}.
\end{align}
All remaining $15$ data points are treated as independent, with the
diagonal covariance $C_{ii} = \sigma_{i}^{2}$.
\section{Analytic growth factor for Type-3 NGR}
\label{app:growth}
In the following, we derive the closed-form solution of the M\'esz\'aros
equation~\eqref{mesroasina} used in Fig.~\ref{fig:growth}.
\subsection{Reduction to a hypergeometric equation}
The M\'esz\'aros equation \eqref{mesroasina} is
\begin{equation}\label{eq:A.1}
\frac{d^2 \delta _m}{da^2}+3\left(\frac{1}{a}-\frac{1}{2a \left(1+a^3 \mathcal{R}_m\right)}\right) \frac{d \delta _m}{da}- \frac{\left(2-c_2\right) \delta _m}{2 a^2 c_2\left(1+a^3 \mathcal{R}_m\right)}=0\,.
\end{equation}
We look for a solution of the form
\begin{equation}
  \delta_m(a) = a^{\alpha}\, u(x), \qquad x \equiv -a^3 \mathcal{R}_m,
  \label{eq:A.2}
\end{equation}
with $\alpha$ to be determined. Since $x \to 0$ as $a \to 0$, the
ansatz isolates the early-time behaviour $\delta \propto a^{\alpha}$,
which is the standard power-law growth in the matter era.

Computing the derivatives,
\begin{align}
  \frac{d \delta _m}{da} &= \alpha a^{\alpha-1} u - 3 \mathcal{R}_m a^{\alpha+2} u', \\
 \frac{d^2 \delta _m}{da^2} &= \alpha(\alpha-1) a^{\alpha-2} u
           - 6(\alpha+1) \mathcal{R}_m a^{\alpha+1} u'
           + 9 \mathcal{R}_m^2 a^{\alpha+4} u'',
\end{align}
where $u' \equiv du/dx$. Substituting into Eq.~\eqref{eq:A.1},
multiplying by $a^{2-\alpha}$, and using $\mathcal{R}_m a^3 = -x$, we obtain
\begin{equation}
9x^2\,u''+ x\left[6\alpha+15-\frac{9}{2(1-x)}\right]u'+\left[\alpha(\alpha-1)+3\alpha-\frac{3\alpha+(2-c_2)/c_2}{2(1-x)}\right]u=0\,.
\label{eq:A.3}
\end{equation}
\subsection{\texorpdfstring{Fixing $\alpha$ and identifying the hypergeometric parameters}{Fixing alpha and identifying the hypergeometric parameters}}

The hypergeometric equation
$x(1-x)u'' + [C - (A+B+1)x]u' - AB\,u = 0$
has no $x^{-1}$ singularity: the coefficient of $u$, after dividing by
$x$, must reduce to a pure constant. Multiplying Eq.~\eqref{eq:A.3}
by $(1-x)/(9x)$, the $u$-coefficient reads
\begin{equation}
  \frac{1}{9x}\Big[\big(\alpha^2+2\alpha\big)(1-x)
  - \tfrac{1}{2}\!\left(3\alpha + \tfrac{2-c_2}{c_2}\right)\Big].
\end{equation}
For this to be independent of $x$, the constant term in the numerator
must vanish:
\begin{equation}
  2\alpha^2 + \alpha - \frac{2-c_2}{c_2} = 0.
  \label{eq:A.4}
\end{equation}
This is precisely the indicial condition at $x=0$. It selects the exponent $\alpha$ for which the growing-mode
solution is regular at early times. Its larger root is
\begin{equation}
  \alpha \equiv \frac{-1 + \sqrt{-7+16/c_2}}{4}.
  \label{eq:A.5}
\end{equation}
The smaller root, $\alpha_- = (-1 - \sqrt{-7+16/c_2})/4$, corresponds
to the decaying mode, which we discard.

With the choice~\eqref{eq:A.5}, the equation becomes
\begin{equation}
  x(1-x)\,u'' +
  \left[\frac{4\alpha+7}{6} - \frac{2\alpha+5}{3}\,x\right] u'
  - \frac{\alpha(\alpha+2)}{9}\,u = 0,
  \label{eq:A.6}
\end{equation}
the standard hypergeometric equation
$x(1-x)u'' + [C-(A+B+1)x]u' - AB\,u = 0$ with
\begin{equation}
  A = \frac{\alpha}{3}, \qquad
  B = \frac{\alpha+2}{3}, \qquad
  C = \frac{4\alpha+7}{6}.
  \label{eq:A.7}
\end{equation}

\subsection{The growing-mode solution}

The solution of Eq.~\eqref{eq:A.6} regular at $x=0$ is
$u = {}_2F_1(A,B;C;x)$. From~\eqref{eq:A.2},
we obtain the growth factor
\begin{equation}
  \delta_m(a;c_2) = \mathcal{A}\, a^{\alpha}\,
  {}_2F_1\!\left(\frac{\alpha}{3},\, \frac{\alpha+2}{3};\,
  \frac{4\alpha+7}{6};\, -a^3 \mathcal{R}_m\right),
  \label{eq:A.8}
\end{equation}
with $\alpha$ given by Eq.~\eqref{eq:A.5} and $\mathcal{A}$
determined by the initial condition
$\delta_m(a_{\rm ini}) = a_{\rm ini}$.

\subsection{The \texorpdfstring{$\Lambda$CDM}{LambdaCDM} limit and the range of validity}

In the $\Lambda$CDM limit $c_2 = 1/2$, Eq.~\eqref{eq:A.5} gives
$\alpha = 1$, and Eq.~\eqref{eq:A.8} reduces to the standard growth
factor \cite{Silveira:1994yq}
\begin{equation}
  D_+(a) = a\,{}_2F_1\!\left(\frac{1}{3},\, 1;\, \frac{11}{6};\,
           -a^3 \mathcal{R}_m\right).
  \label{eq:A.9}
\end{equation}
For $\Omega_{m0}=0.3$ and $a=1$, this gives $D_+(1)\approx0.78$. The cosmological constant suppresses growth relative to the Einstein--de Sitter scaling $\delta\propto a$. For the growing root, $\alpha\ge-1/4$ for every $c_2\in(0,16/7]$, so that $C=(4\alpha+7)/6\ge1$ is never a non-positive integer and the series is well defined. The coincidences $A=C$ or $B=C$ would require $\alpha=-7/2$ or $\alpha=-3/2$, which are values of the decaying root $\alpha_-$ (at $c_2=1/11$ and $c_2=1/2$, respectively) and never occur on the growing branch and $A=B$ is impossible for any real $\alpha$. The parametrization \eqref{eq:A.8} is therefore nondegenerate for all $c_2\in(0,16/7]$. For $0.2<c_2<0.5$ one finds $\alpha>1$, and the growth rate is enhanced relative to $\Lambda$CDM, as expected for $G_{\rm eff}>G$.

\section*{Acknowledgments} BM acknowledges the support of Anusandhan National Research Foundation(ANRF) for the project(File No: CRG/2023/000475). SKM and BM acknowledge IUCAA, Pune (India) for providing support in the form of an academic visit during which part of this work is accomplished. 
\bibliographystyle{JHEP}
\bibliography{references}
\end{document}